\documentclass[conference]{IEEEtran}
\IEEEoverridecommandlockouts
\usepackage{cite}
\usepackage{graphicx,graphics,epsfig,subfigure,times,bm,bbm,amssymb,amsmath,amsfonts,amsthm,mathrsfs,MnSymbol,xcolor}
\usepackage{gensymb}
\usepackage[matrix,frame,arrow]{xypic}
\usepackage[pdfstartview=FitH]{hyperref}
\usepackage{subfigure}
\usepackage{braket}
\usepackage{enumerate}
\usepackage{tikz}
\usetikzlibrary{calc}
\usepackage{float}

\usepackage{amsmath,amssymb,amsfonts}
\usepackage{algorithmic}
\usepackage{graphicx}
\usepackage{textcomp}
\usepackage{xcolor}
\def\BibTeX{{\rm B\kern-.05em{\sc i\kern-.025em b}\kern-.08em
    T\kern-.1667em\lower.7ex\hbox{E}\kern-.125emX}}
\begin{document}

\title{Optimal resource requirements for connected quantum sub-networks\\
}

\author{
\IEEEauthorblockN{Shashank Shekhar}
\IEEEauthorblockA{\textit{Department of Physics and QuICST} \\
\textit{Indian Institute of Technology Bombay}\\
Mumbai 400076, India \\
shashankshekhar01234@gmail.com}
\and
\IEEEauthorblockN{Md Sohel Mondal}
\IEEEauthorblockA{\textit{Department of Physics and QuICST} \\
\textit{Indian Institute of Technology Bombay}\\
Mumbai 400076, India \\
mondalsohelmd@gmail.com}
\and
\IEEEauthorblockN{Siddhartha Santra}
\IEEEauthorblockA{\textit{Department of Physics and QuICST} \\
\textit{Indian Institute of Technology Bombay}\\
Mumbai 400076, India\\
santra@iitb.ac.in}
}
\maketitle
\begin{abstract}
The realization of a global quantum network capable of supporting secure communication and other quantum information processing (QIP) tasks hinges on the ability to distribute high-fidelity entanglement across long distances while optimizing resource usage. This work describes a scalable approach for building large quantum networks by connecting quantum sub-networks using entanglement backbones as interconnections and a swapping based entanglement distribution protocol. Using a statistical model for parametrized quantum sub-networks we derive a set of equations whose solutions give the optimal values of average network parameters that meet threshold requirements for QIP tasks while minimizing resource cost functions. Our analysis extends to the scenario where multiple sub-networks must be interconnected simultaneously based on the formulation of a global resource cost function. The probability of successfully satisfying the parameter thresholds of a QIP task as a function of average parameters of the sub-networks for random network demands reveals a transition from zero to full satisfiability above critical network parameter values. Moreover, we find that the satisfiability transition can be smooth or discontinuous depending on the topology of the sub-networks. Our results present a pathway for calculating optimal resource requirements in quantum sub-networks interconnected to form the global quantum internet.
\end{abstract}
\begin{IEEEkeywords}
quantum sub-networks, entanglement swapping, quantum resource optimisation.
\end{IEEEkeywords}
\section{Introduction}

The development of a global quantum network \cite{qn_pirandola,qn_wehner,qn_simon} marks a significant leap in quantum information science, offering transformative capabilities for secure communication, distributed quantum computing \cite{Komar2014,Yin2020}, and advanced quantum information processing (QIP) tasks. Quantum networks leverage non-local resources, such as entanglement, to transmit quantum information across distant network nodes. However, it is experimentally challenging to build large networks that can preserve entanglement between geographically separated nodes while mitigating the effects of decoherence in the quantum channels and memories.

\begin{figure}
    \centering
    \includegraphics[width=\linewidth]{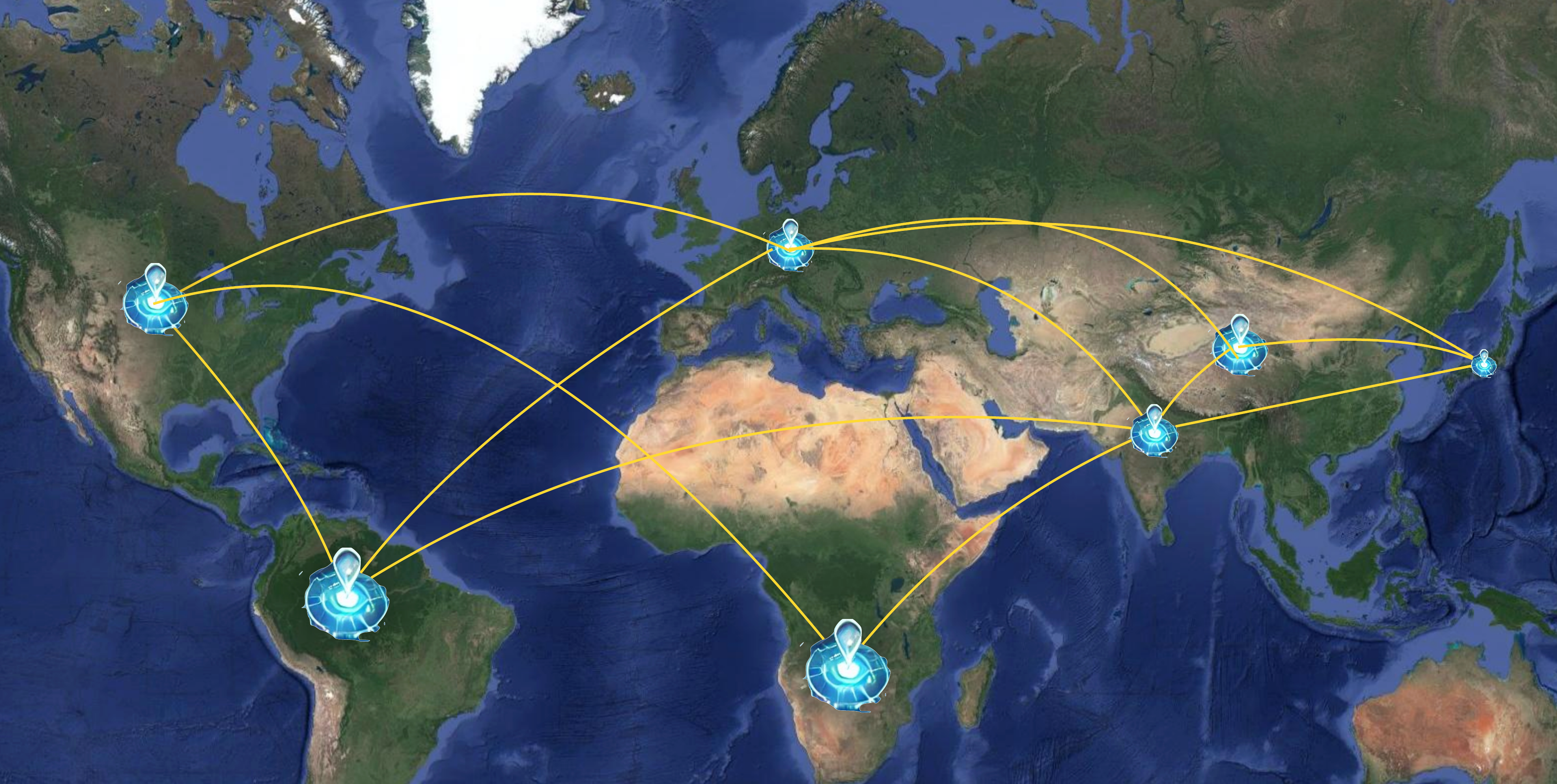}
    \caption{Visualization of a global quantum network model showing regional quantum sub-networks represented by blue symbols distributed across various geographical locations around the world. These sub-networks are interconnected via entanglement backbones (yellow links) which are high-fidelity, high-bandwidth quantum channels. The image was created using the Google Maps.}
    \label{global_network}
\end{figure}

A promising strategy for constructing a global quantum network involves the deployment of smaller \textit{quantum sub-networks} which can achieve high fidelity entanglement distribution within localized regions \cite{Yehia_2024,Chen2021}. These sub-networks, connected via entanglement backbones which are long-distance, high-bandwidth connections offer a way to achieve scalabality of the overall network size, see Fig. \ref{global_network}. Reducing the resource cost of constructing these subnetworks, in terms of the average entanglement over the subnetwork edges, can yield feasible target experimental benchmarks - thereby advancing the practical realization of large-scale quantum networks.

\emph{Our contribution}:- In this work we present an approach for optimizing the resource cost of creating connected quantum sub-networks under the constraint of meeting the entanglement threshold requirement for a given but arbitrary QIP task. We elucidate our optimisation approach using a simple entanglement distribution protocol based only on entanglement swapping of adjacent entangled edges of the quantum sub-networks. Our calculations are based on a statistical model for quantum networks \cite{mondal_ent_topography} that captures the stochasticity of the entanglement generation process over the network edges using a random variable characterisation of two edge parameters: the fidelity of the state over an edge and the probability of successful entanglement generation over the edge. 

We formulate a resource cost for the entire set of connected sub-networks that is extensive in their number and the sizes of these networks. The resource cost is a non-linear function of the average edge-fidelity and average edge-probability of the sub-networks to account for the resources consumed in increasing the average edge-fidelity (using entanglement purification \cite{purification_benett},\cite{purification_deutsch}) and average edge-probability over the network edges (using multiplexing \cite{multiplexing_PhysRevA}). The inequality constraint coming from the threshold requirement is also non-linear due to the form of the end-to-end fidelity achieved along a network path using the swap-based distribution protocol. We solve this constrained non-linear optimisation problem using the Lagrangian method in the parameter regions where the Karush–Kuhn–Tucker (KKT) conditions \cite{boyd2004convex} are satisfied. The solution of the optimisation problem yields a set of non-linear equations which in turn are solved numerically to obtain optimised values of average edge-fidelities and edge-probabilities for each of the participating sub-networks that can ensure reliable execution of the QIP task throughout the entire connected network. 

Further, we simulated a set of demands, for connecting a pair of nodes, on specific configurations of connected sub-networks and verifying if the demands are satisfied for a QIP task, such as QKD \cite{Diamanti2016,QKD,Stanley_2022}, as we scan through different values of the average network parameters. We find that above some certain values of the network parameters most network demands can be met while for average network parameter values below the that values random network demands are rarely satisfied.


The rest of the paper has the following structure: Subsec.~\ref{sec_two_subsec_one}
describes quantum sub-networks while the statistical model of quantum networks used in this work is described in Subsec~\ref{sec_two_subsec_two}. The end-to-end parameter values obtained on network paths via the entanglement swapping based distribution protocol is described in Subsec.~\ref{sec_two_subsec_three}. A derivation of the inequality constraint arising from requirements of threshold satisfiability using two connected sub-networks is described in Subsec.~\ref{sec_three_subsec_one}. Next, we formulate the resource cost function in Subsec.~\ref{sec_three_subsec_two}. Following on, we present the constrained optimisation procedure of the resource cost function in Subsec.~\ref{sec_three_subsec_three}. Then, in Sec.~\ref{sec_four} we provide numerical results that capture the satisfiability of random network demands for various connected sub-network configurations as functions of average network-parameter values. Finally, we
conclude with a discussion and scope for future work in Sec.~\ref{sec_five}.

\section{Background: Quantum sub-networks, Entanglement backbone, Statistical network model, Swapping-based entanglement distribution protocol}
\label{sec_two}
In this section we describe the preliminaries required for our work. Specifically, we describe quantum sub-networks and entanglement backbones followed by the statistical network model we use. Then we discuss the swapping-based entanglement distribution protocol and the calculation of random path parameter values, viz. path-fidelity and path-probability, using our statistical network model. 

\begin{figure}
    \centering
    \includegraphics[width=\linewidth]{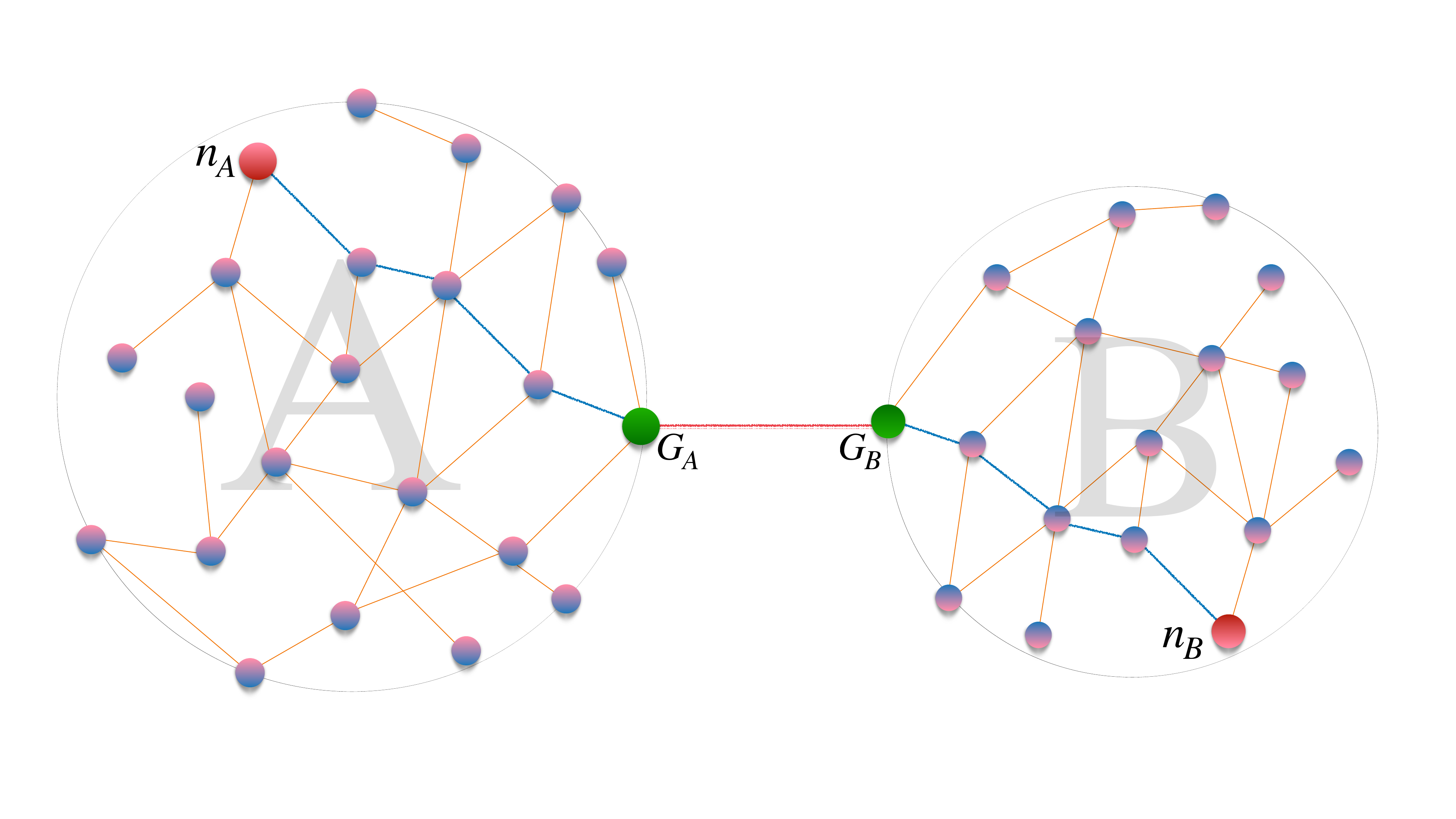}
    \caption{$A$ and $B$ are two smaller sub-networks of a global quantum network. $G_A$ and $G_B$, shown in green, are the gateways of networks $A$ and $B$ respectively which are connected via a link (shown in red line). $n_A$($n_B$), shown in red, is the furthest node from the gateway $G_A$($G_B$) in the sub-network $A$($B$). We call $n_A$ and $n_B$ as edge nodes of $A$ and $B$ respectively. The shortest graph path (SGP) between $n_A$($n_B$) and $G_A$($G_B$) is shown in blue lines. Whenever a node $n_A$ in $A$ wants to make a connection to any node $n_B$ in $B$, first $n_A$($n_B$) will be connected to $G_A$($G_B$) via entanglement swapping at all the intermediate nodes along the SGP and then two entanglement swappings at the gateways $G_A$ and $G_B$ will accomplish a connection between $n_A$ and $n_B$.}
    \label{sub_networks}
\end{figure}

\subsection{Quantum sub-networks}
\label{sec_two_subsec_one}
A quantum sub-network is essentially a small quantum network spanning a relatively localized geographic region compared to the size of the overall network \cite{Yehia_2024}. The classical analogues being metropolitan area telecom networks \cite{metropolitan_net} and internet local-area networks \cite{local_area_net} or wide-area networks \cite{wide_area_net}. The limited distances between the nodes within a quantum sub-network allow for a higher degree of connectivity among its nodes - and therefore, the nodes within a sub-network are tightly clustered \cite{complex_net}. Moreover, the shortest graph path distances between any two nodes within a sub-network are small with only a few hops required along such paths, see Fig. \ref{sub_networks}. 

In terms of geographic distance, $d_{ij}$, between the nodes of a network, one can identify a sub-network, $A$, as a distinguished set of nodes, $\{i|i\in A\}$ along with the edges between such nodes, which are connected in the graph theoretic sense \cite{steen_complex_networks} and whose average geographic separation is much smaller than any inter-network distance in the overall set of connected sub-networks. That is,
\begin{align}
    \langle d^{\text{intra}}(i,j) \rangle_{i,j\in A}, \langle d^{\text{intra}}(i,j) \rangle_{i,j\in B}&\ll \langle d^{\text{inter}}(i,j) \rangle_{i\in A, j\in B},
    \label{sub_network_defn}
\end{align}
where $\langle d^{\text{intra}}(i,j) \rangle_{i,j\in A}$ and $\langle d^{\text{intra}}(i,j) \rangle_{i,j\in B}$ are the average geographic distances between nodes within a sub-network $A$ and $B$ respectively, and $\langle d^{\text{inter}}(i,j) \rangle_{i\in A, j\in B}$ is the geographic distance between any two nodes across different sub-networks.

Operationally, within a quantum sub-network complex entanglement manipulation operations such as purification over an edge to increase the fidelity of the entangled state \cite{deutsch,Bennett}, or multiplexing to increase the effective probability of entanglement generation \cite{multiplexing_PhysRevA} are deemed more feasible than across sub-networks. This stems from their smaller geographical scale resulting in shorter communication distances and correspondingly lower decoherence. As a result, operations like entanglement purification which require high initial input fidelities for successful purification are significantly more effective within a sub-network compared to long-distance operations across sub-networks that are more complex but less useful. Effectively, a quantum sub-network allows greater control, more sophistication and higher efficiency of entanglement manipulation operations and can therefore serve as a building block of a larger quantum network obtained by inter-connecting multiple sub-networks.
 
\subsection{Entanglement backbone for connecting sub-networks}

Entanglement backbones are physical channels that allow high-fidelity entangled quantum states to be shared with high-rates between any two quantum sub-networks. These backbones may be practically realized through various media such as satellite links \cite{Scott_Bloom:03}, fiber-optic cables \cite{gisin_crypto} or vacuum beam guides \cite{vacuumbeamguidelargescale} - each of which has its own advantages. Although, entanglement backbones can be physically distinguished by the share of physical hardware resources allocated to them (higher-bandwidth channels, superior quantum hardware resources) - mathematically, they can be well described as a single-edge of the global network albeit with a high-fidelity of the generated quantum state as well as high-probability of entanglement generation over the backbone.

Interconnecting sub-networks via entanglement backbones provides a scalable framework for a global quantum network. Each sub-network has a distinguished node (or set of nodes), called as gateways, that interface with the entanglement backbone for communication with external sub-networks, see Fig. \ref{sub_networks}. Any node within a particular sub-network can communicate with a node within an external sub-network by first connecting with its own sub-network gateway. In this way, entanglement backbones are useful to obtain a modular scheme for scaling quantum networks that leverage reliable and robust entanglement manipulations over individual sub-networks while enabling long-distance quantum communication.

\subsection{Statistical network model}
\label{sec_two_subsec_two}

Any quantum network (therefore also the sub-networks) can be mathematically modeled based on an underlying graph, $G(V,E)$. The set of network nodes, $V$, represents stationary quantum systems (qubits or quantum memories), while the set of edges, $E=\{(v_i,v_j)|i,j\in V\}$, denotes the physical links that enable direct classical and quantum communication between a subset of nodes, $i,j\in V$, including the possibility of entanglement generation between adjacent nodes. 

The physical connections represented by the edges may comprise of fiber-optic channels, vacuum waveguides, or satellite-based free-space channels and exhibit distinct physical properties depending on the underlying technology employed. Within quantum sub-networks fiber-optic channels are more feasible while connections across sub-networks are likely to be achieved using the other types of physical channels mentioned above.

The fundamental process of entanglement generation over the edges of a quantum network is probabilistic and results in a quantum state, $\rho(F_e)$, over the edge with,
\begin{align}
    \rho(F_e)=\frac{4F_e-1}{3}\ket{\phi^+}\bra{\phi^+} + \frac{1-F_e}{3}\bold{1}_4,
    \label{isotropic-state-fidelity}
\end{align}
which has a fidelity, $\braket{\phi^+|\rho(F_e)|\phi^+}=F_e$, with respect to the maximally entangled two-qubit state, $\ket{\phi^+} = (\ket{00}+\ket{11})/\sqrt{2}$.

The two quantities, the fidelity of the state over an edge ($F_e$) and the probability of succesful entanglement generation ($\eta_e$) can be modeled as random variables with their respective probability distributions  $p_F(F_e), 0\leq F_e\leq 1$ and $p_\eta(\eta), 0\leq \eta_e\leq 1$. A high value of the average edge-fidelity $\bar{F_e}\to 1$ over a network reflects that most edges have low decoherence and high-quality entangled states may be established over them. A high value of the average network edge-probability $\bar{\eta_e}\to 1$ can describe the situation where the channels of the network have low loss. 

Together, the knowledge of the network graph, the form of the edge states shown in Eq. (\ref{isotropic-state-fidelity}) and the edge-parameter distributions in the form of the tuple,
\begin{align}
    Q := \{ G(V,E), \rho(F_e), p_F(F_e), p_\eta(\eta_e)\},
    \label{network}
\end{align}
fully defines the statistical model of quantum networks used in this work. 

\subsection{Swapping-based entanglement distribution protocol}
\label{sec_two_subsec_three}

The entanglement generated along adjacent native edges of the network graph $G(V,E)$ can be connected together to obtain long-range entangled states in a quantum network. A simple protocol to achieve this is via entanglement swapping \cite{swapping_santra,avgconc_santra} at intermediate nodes between the adjacent entangled edges along a network path. Entanglement swapping is a  local quantum operation aided by classical communication (LOCC) that is essential for establishing multi-hop entanglement over distant network nodes at the cost of reduced fidelity and reduced probability with increasing number of hops.

For a network path comprised of $l$-adjacent edges, the fidelity of the entangled state between the terminal nodes of the path and the probability of obtaining this state via entanglement swapping at intermediate nodes is also a random variable - since the path parameters depend on the fidelities, $F_e$, and probabilities, $\eta_e$, for edges along the path. Assuming independent and identically distributed (i.i.d.) edge fidelities and probabilities along any path, the average fidelity and probability (over all paths of length $l$) of the final state after \((l-1)\)-swaps is,
\begin{align}
    F_l &= \frac{1}{4} + \frac{3}{4} \left(\frac{4\overline{F}_e - 1}{3}\right)^l,\nonumber\\
    \eta_{l} &= \left( \overline{\eta}_e \right)^l,
    \label{l_swap_fidelity}
\end{align}
where, $\overline{F}_e$ is the average edge fidelity in the network. These equations show the dependence of average path parameters on the path length as well as the statistical mean of edge-parameters in a given quantum network. 

\section{Path parameters for connected quantum sub-networks}
\label{sec_three}
 Here we consider the path parameters arising out of inter-sub-network connections when the entanglement backbone is taken as an intermediate edge along the path and obtain inequalities arising out of threshold satisfaction constraints. Then we describe formulation of the resource cost functions and finally we present the formal constrained non-linear optimisation problem formulation.

\subsection{Path parameters for inter-network connections}
\label{sec_three_subsec_one}
To ensure that any two nodes in the global quantum network can be connected for a given QIP task, each sub-network must meet specific requirements for the edge parameters. These required values will depend on the state that connects the sub-networks and the thresholds for the given QIP task. The analysis begins by deriving an inequality for average fidelity in sub-network, which establishes the minimum average fidelity required for secure and effective entanglement distribution between any two nodes across the sub-network. Further, the next step is to derive a corresponding inequality for average probability. The calculations are initially demonstrated for the case of two quantum sub-networks and then generalized for 

Consider a scenario within our two-sub-network model as shown in Fig.~\ref{sub_networks}. We want to establish communication and perform a specific QIP task between a chosen node, denoted as $n_A$, within sub-network $A$ and another node, $n_B$, residing in sub-network $B$. Each sub-network has its own dedicated gateway, $G_A$ and $G_B$ respectively which basically connects the sub-networks with the backbone. These gateways play a critical role in enabling communication across the larger network. Crucially, $G_A$ and $G_B$ are entangled with each other, but not necessarily in a perfect Bell state. Instead, we can represent their entanglement using a more general isotropic state with a specific fidelity denoted by $F^*$ and probability denoted by $\eta^*$.

Our objective here is to leverage entanglement swapping to connect $n_A$ and $n_B$ and subsequently perform a specific QIP task between them. For this we connect $n_A$ with $G_A$ and $n_B$ with $G_B$ by performing $l_1$ and $l_2$ swaps respectively. The fidelity of the resulting entangled links are denoted by $F_{l_1}$ and $F_{l_1}$, given by Eq. (\ref{l_swap_fidelity}) and the multiplexed probability of the resulting entangled links are denoted by $\eta_{l_1}$ and $\eta_{l_2}$, given by Eq. (\ref{l_swap_eta_l}). Then we perform the last two swaps on node $G_A$ and $G_B$ to finally connect the $n_A$ and $n_B$. The resulting fidelity $F_{n_A n_B}$ is given by: 
\begin{align}
    F_{n_A n_B} = \frac{1}{9}( &16F_{l_1}F_{l_2}F^* - 4F_{l_1}F_{l_2} - 4F^*F_{l_1} \nonumber \\
    &- 4F_{l_2}F^* +  F_{l_1} + F_{l_2} + F^* + 2),
\end{align}
and the resulting $\eta_{n_A n_B}$ is given by: 
\begin{align}
    \eta_{n_A n_B} = \eta_{l_1}\eta^* \eta_{l_2}.
\end{align}
The parameters $F_{n_A n_B}$ and $\eta_{n_A n_B}$ needs to above a certain threshold for any given QIP task: $F_{n_A n_B} \geq F_{Th}$ and $\eta_{n_A n_B} \geq \eta_{Th}$. However, the average parameters of each sub-network becomes a critical factor as $F_{n_A n_B}$ and $\eta_{n_A n_B}$ both depends on the average edge parameters of both the sub-networks, $\overline{F}_e$ and $\overline{\eta}$. This gives us the required inequalities for fidelity and probability. The inequality for fidelity can be expressed as:
\begin{align}
    g(\overline{F}_{e_1},\overline{F}_{e_2}) = f_{eff} -  \left(4F_{l_1} F_{l_2} - F_{l_1} - F_{l_2}\right) \leq 0,
    \label{Fidelity_inequality}
\end{align}
where,
\begin{align}
     f_{eff} = \frac{9F_{Th} - F^* - 2}{4F^* -1}, 
     \label{f_eff}
\end{align}
and $F_{l_1}$ and $F_{l_2}$ is of the form given by Eq.~(\ref{l_swap_fidelity}).
 
Similarly, the inequality for probability can be expressed as:
\begin{align}
    h(\overline{\eta}_{e_1}, \overline{\eta}_{e_1}) = \eta_{\text{Th}} - \eta_{l_1} \eta^* \eta_{l_2} \leq 0,
    \label{eta_inequality}
\end{align}
where, $\eta_{l_i}$ is of the form given by Eq.~(\ref{l_swap_eta_l}).

In a realistic quantum internet, multiple sub-networks are expected to be distributed across the globe. For certain quantum information processing (QIP) tasks, it may be necessary to simultaneously connect more than two sub-networks. Ensuring that both the fidelity constraints, given by Eq.~(\ref{Fidelity_inequality}), and the probability constraints, given by Eq.~(\ref{eta_inequality}), are satisfied for all possible pairs of sub-networks is the requirement in such case. This imposes a global constraint on the network’s performance, ensuring successful execution of QIP tasks (such as quantum key distribution, QKD) between any two nodes located in different sub-networks. Thus the set of  constraints for a global network setup can be formalized as: $g(\overline{F}_{e_i}, \overline{F}_{e_j}) \leq 0 ~\text{and}~ h(\overline{\eta}_{e_1}, \overline{\eta}_{e_j}) \leq 0 ~\{\forall{i,j} \in \text{N}:~j>i\}$ where N is the total number of sub-networks.

\subsection{Resource cost functions}
\label{sec_three_subsec_two}
The constraint inequalities given in Eq.~(\ref{Fidelity_inequality}) and Eq.~(\ref{eta_inequality}) defines the feasible region for average edge parameters (\(\overline{F}_{e_1}\), \(\overline{F}_{e_2}\), \(\overline{\eta}_{e_1}\) and \(\overline{\eta}_{e_2}\)), required for each sub-network to successfully execute the given QIP task. However, meeting these thresholds alone is not enough; the resource costs required to establish these conditions must also be optimized. We define cost functions for fidelity and entanglement generation probability distribution across the network.
Among alternate possible cost functions we consider the two defined in Eq.~(\ref{Nonlinear_cost}) and Eq.~(\ref{Rescource_cost_function}) corresponding to fidelity and probability respectively. These cost functions incorporate the resource requirements for feasibility of quantum network protocols such as entanglement purification and multiplexing used to achieve the desired value of average edge parameters. The cost functions are also extensive with the total number of sub-networks as well as with the total number of edges in each sub-network. 

To achieve the required average fidelity on each edge we need to use entanglement purification protocol. Then the cost associated with this can be defined in terms of the number of copies required as target state to achieve purification with non zero yield. For one sub-network this cost associated to fidelity goes as:
\begin{align}
    C_{F}=\frac{|E|}{\sqrt{1 - \overline{F}_{e}}},
    \label{Nonlinear_cost}
\end{align}
where, $|E|$ is the total number of edges in the sub-network. Similarly, to achieve desired probability in the edges we need to do multiplexing. 
The multiplexed probability for an edge of length \(L\) and \(n\) attempts can be written as:
\begin{align}
    \eta(L, n) = 1 - (1 - \eta^{bare}(L))^n,
\end{align}
where \(\eta^{bare}(L)\) is the single-attempt success probability. Assuming i.i.d. edge parameters, the total multiplexed probability after performing \((l-1)\)-swaps across \(l\) edges simplifies to:
\begin{align}
    \eta_l(n, l) = \left(1 - (1 - \overline{\eta}^{bare}_e)^n \right)^l = \left( \overline{\eta}_e \right)^l,
    \label{l_swap_eta_l}
\end{align}
where, \(\overline{\eta}_e\) is the average multiplexed probability. Thus, \(\eta_l(n, l)\) explicitly depends only on the number of attempts \(n\) and edges \(l\), if \(\overline{\eta}^{bare}_e\) is assumed to be a constant for the given network. In this scenario, the resource cost is defined based on the average multiplexed probability, $\overline{\eta}_e$, which itself depends on the average generation probability, $\overline{\eta}^{bare}_e$, and the number of attempts, $n$, per block during multiplexing, as described in Eq.~(\ref{l_swap_eta_l}). The resource cost function can be expressed as:
\begin{align}
    C_{\eta} = C |E| n,
\end{align}
where $C$ represents a constant resource cost associated with each edge which can be taken as 1. These values can be formulated in terms of the multiplexed and bare average probabilities of the network. Consequently, the cost function becomes:
\begin{align}
     C_{\eta} = |E| \left(\frac{\log (1-\overline{\eta}_{e})}{\log (1-\overline{\eta}^{bare}_{e})} \right).
    \label{Rescource_cost_function}
\end{align}
So, the global entanglement cost function, capturing the total resource expenditure for entanglement establishment, can be expressed as the summation of individual entanglement costs across all sub-networks, $\sum_i C_{F_i}$. Similarly, the global resource cost for multiplexing is the summation of individual resource costs for each sub-network, $\sum_i C_{\eta_i}$.

\subsection{Constrained non-linear optimization of resource cost functions}
\label{sec_three_subsec_three}
In this subsection we focus on optimizing the two independent cost functions associated with fidelity and edge probability defined in Eq.~(\ref{Nonlinear_cost}) and Eq.~(\ref{Rescource_cost_function}) respectively satisfying the constraints given in Eq.~(\ref{Fidelity_inequality}) and Eq.~(\ref{eta_inequality}). This boils down to solving two independent optimization problems related to these two edge parameters respectively. We use the Karush-Kuhn-Tucker (KKT) method of optimization \cite{boyd2004convex} which serves our purpose as we have an inequality as constraint. In KKT method it is standard to convert a primal optimization problem into a dual optimization problem, which involves defining a Lagrangian and dual variables. We formulate the Lagrangians for the optimization of the two cost functions as given below,
\begin{align}
    L_F = \sum_i \left(C_{F_i} +  \sum_{j > i} \lambda_{ij}g(\overline{F}_{e_i}, \overline{F}_{e_j}) \right),
    \label{n_network_lagrangian}
\end{align}
\begin{align}
    L_{\eta} = \sum_i \left(C_{\eta_i} +  \sum_{j > i} \mu_{ij} h(\overline{\eta}_{e_i}, \overline{\eta}_{e_j}) \right).
    \label{eta_n_network_lagrangian}
\end{align}
Here, $\lambda_{ij}$ and $\mu_{ij}$ are the Lagrange multipliers for the two Lagrangians. We consider a simpler case of two sub-networks to obtain Eq.~(\ref{Fidelity_sol_1}) and Eq.~(\ref{eta_sol_1}). These set of equations can further be solved numerically to yield the optimized average edge fidelities and average edge probabilities of the sub-networks.
\begin{align}
    \frac{|E_1|(4\overline{F}_{e_1}-1)}{l^1_{max}(1-\overline{F}_{e_1})^{3/2}} = \frac{|E_2|(4\overline{F}_{e_2}-1)}{l^2_{max}(1-\overline{F}_{e_2})^{3/2}},\nonumber
\end{align}
\begin{align}
    \frac{9}{4}\left(\frac{(4\overline{F}_{e_1}-1)}{3}\right)^{l^1_{max}}\left(\frac{(4\overline{F}_{e_2}-1)}{3}\right)^{l^2_{max}} + \left(\frac{1}{4}+f^{eff}\right) = 0,
    \label{Fidelity_sol_1}
\end{align}
\begin{align}
    \frac{|E_1|(\overline{\eta}_{e_1})}{l^1_{max}(1-\overline{\eta}_{e_1})} = \frac{|E_2|(\overline{\eta}_{e_2})}{l^2_{max}(1-\overline{\eta}_{e_2})},\nonumber
\end{align}
\begin{align}
    (\overline{\eta}_{e_1})^{l^1_{max}} \eta^* (\overline{\eta}_{e_2})^{l^2_{max}}  = \eta_{Th}.
    \label{eta_sol_1}
\end{align}
Solving these pair of equations numerically, we obtain the optimal average edge parameter values for the sub-networks - $\overline{F}_{e_1}$, $\overline{F}_{e_2}$, $\overline{\eta}_{e_1}$ and $\overline{\eta}_{e_2}$ as functions of network parameters $|E_1|$, $|E_2|$, $l^1_{max}$ and $l^2_{max}$. These optimization approach provides a holistic understanding of resource allocation strategies in quantum networks. By optimizing both fidelity and probability, secure and efficient quantum information processing can be achieved.

In a general setting of $n$ sub-networks, we get the optimized average edge parameters by solving both Lagrangian given in Eq.~(\ref{n_network_lagrangian}) and Eq.~(\ref{eta_n_network_lagrangian}) and minimizing them with respect to the relevant parameters. Thus, we can determine the optimal edge parameters that minimize the total cost for the global quantum internet. The final solution must lie within the feasible parameter range, and among all feasible solutions, the one that minimizes the global cost will yield the desired values for the edge parameters.

\section{Numerical Results: Satisfiability transition for random network demands}
\label{sec_four}
\begin{figure*}[t]
    \centering
    \includegraphics[width=\linewidth]{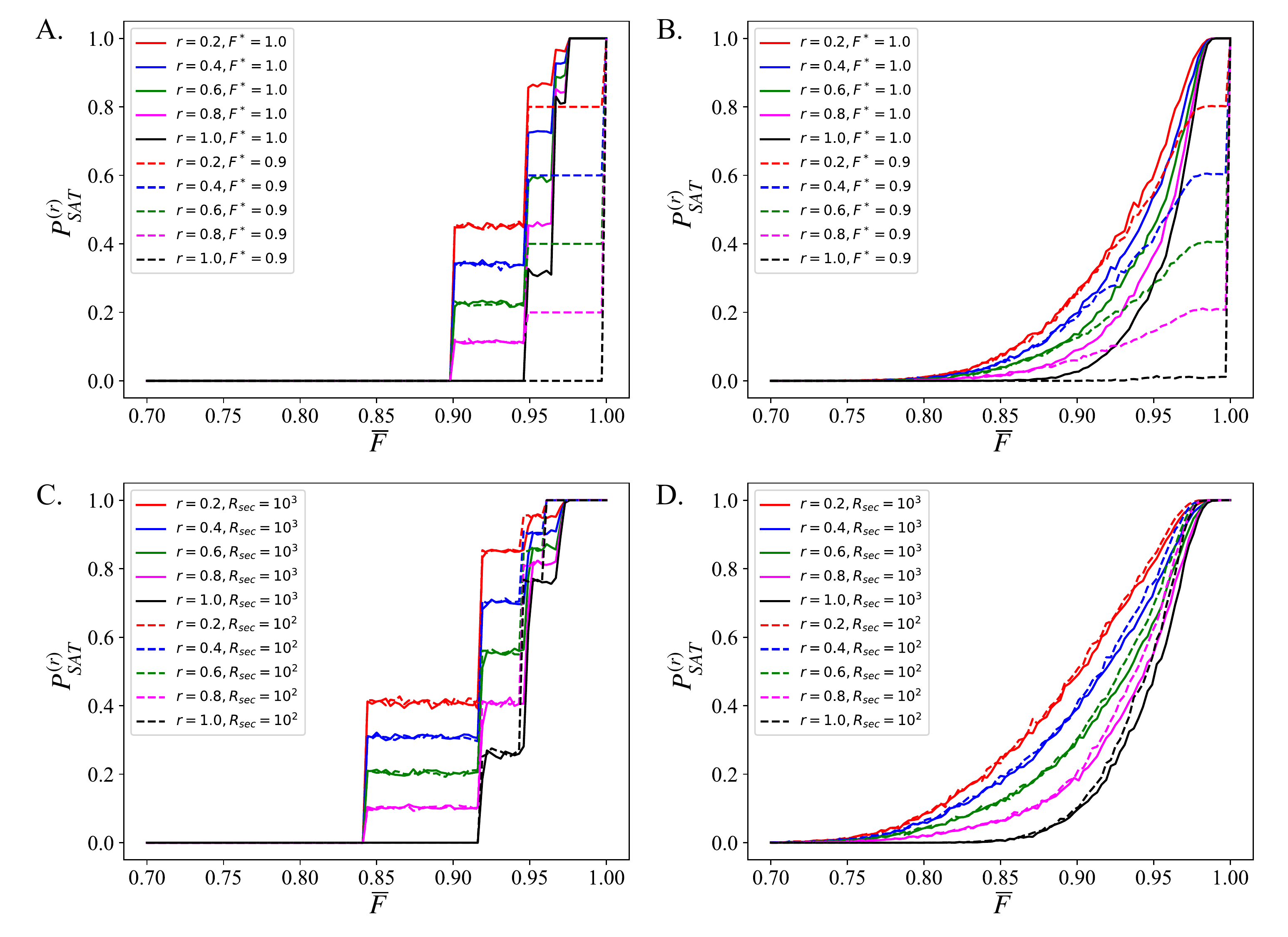}
    \caption{Transition in the probability of satisfiability, Eq. (\ref{eq:psat}), as a function of average edge fidelity across two quantum sub-networks. Panels (A) and (C) correspond to homogeneous networks, while (B) and (D) represent networks with a standard deviation of 0.05 in the edge parameters. (A) and (B) evaluate this probability based on satisfying QKD thresholds (fidelity = 0.9, probability = 1.6\%, with \( R_{\text{eps}} = 10^6 \) and \( R_{\text{sec}} = 10^3 \)) for identical-sized networks. In (C) and (D), the probability is assessed based on achieving the required secure key rate, with networks of distinct sizes. Steep transitions and multiple jumps are observed in homogeneous networks, whereas non-homogeneous networks exhibit smoother transitions. Jumps occur as intra-network demands are first met, followed by inter-network demands for closer nodes. As the inter-network demand ratio \( r \) increases, probability decreases, highlighting the importance of high average fidelity for long-distant reliable quantum communication. In panels A and B the network sizes used for simulation are, $|\text{V}_A|$ = $|\text{V}_B|$ = 60 and $|\text{E}_A|$ = $|\text{E}_B|$ = 1000 where as in panels C and D the network sizes used for simulation are, $|\text{V}_A|$ = 60, $|\text{E}_A|$ = 1000 and $|\text{V}_B|$ = 30, $|\text{E}_B|$ = 200.
}
    \label{phase_transition_plot}
\end{figure*}
In this section, we numerically investigate the probability of satisfying random network demands for different average edge-parameter values of specific configurations of connected quantum sub-networks. For our numerical simulations, we consider a system comprising two quantum sub-networks as shown in Fig.~\ref{sub_networks}. We show the plots for different value of \textit{inter-network demands} and for different scenarios like homogeneous and heterogeneous networks. 

\subsection{Probability of Satisfiability and Ratio of Inter-Network Demands}
\label{sec_four_subsec_one}
The probability of satisfiability refers to the probability with which a given quantum information processing (QIP) task, such as quantum key distribution (QKD) or entanglement-based quantum communication, can be executed successfully between two randomly selected nodes in the global quantum network. 
The probability is computed statistically by sampling multiple pairs of nodes across the global network and evaluating the success rate for different configurations of fidelity and probability. The results are averaged over these random configurations to account for variations in the underlying distribution of edge fidelities, where for a particular configuration, the probability of satisfiability can be defined as,
\begin{align}
    P_{SAT}^{(r)}=\frac{\#D_{satisfied}}{\#D_{total}},
    \label{eq:psat}
\end{align}
where, $\#D_{satisfied}$ is the number of demands that get satisfied based on whether the edge parameter requirements are met or not and $\#D_{total}$ is total number of demands. We define the demand, $D_{S,D}$, between a pair of source-destination nodes in the network quantitatively as a connection request with the threshold values of the pair of the fidelity and probability specified. That is,
\begin{align}
D_{S,D}:=(f_{\rm Th}, \eta_{\rm Th}),
\end{align}
with the demand $D_{S,D}$ being an intra-network demand when the nodes $S,D$ are both nodes in the same sub-network, whereas, it is an inter-network demand when $S,D$ belong to different sub-networks.

Further, a key parameter in this analysis is the \textit{ratio of inter-network demands}, which quantifies the proportion of total communication demands that require inter-network communication. This ratio ($r$) reflects the fraction of quantum communication tasks that span across the two sub-networks. It is defined as: $
r = D_{inter}/(D_{inter} + D_{intra})$, where \( D_{inter} \) represents the number of communication demands between nodes from different sub-networks, and \( D_{intra} \) corresponds to the number of demands within the same sub-network.

As the ratio of inter-network demands increases, the probability is expected to decline. This is due to the fact that inter-network communication involves longer distances, leading to increased decoherence and signal loss, which reduce both fidelity and entanglement generation probability. Our numerical simulations confirm this trend, with higher proportions of inter-network demands resulting in a lower success rate.

\subsection{Secure key rate}
\label{sec_four_subsec_two}
In addition to evaluating the individual thresholds for fidelity and probability, we also analyze the case where one can define the threshold in terms of Secure key rate. The secure key rate, \( R_{\text{sec}} \), represents the achievable rate of generating secure cryptographic keys, given the final fidelity and probability of the quantum channel. It is defined as:
\begin{align}
R_{\text{sec}} = R_{\text{eps}} \cdot P_{\text{path}} \cdot f_{key}(F),
\end{align}
where \( R_{\text{eps}} \) is the repetition rate of entangled pairs in the network, \( P_{\text{path}} \) is the path success probability (the probability that a working path exists between the two nodes), and \( f_{key}(F) \) is the secure key fraction which is a function of the final fidelity \( F \) of the entangled state. The function \( f_{key}(F) \) is given by, $f_{key}(F) = 1 - 2h(1 - F)$. Here, \( h(x) \) is the binary entropy function, defined as, $h(x) = -x \log_2(x) - (1 - x) \log_2(1 - x)$.

This formula reflects the practical requirements for secure quantum communication, where higher fidelities lead to better key generation performance. In our analysis, we compute the probability based on whether the secure key rate between two randomly chosen nodes exceeds the required threshold for secure communication.

\subsection{Transition in the probability of satisfiability}
\label{sec_four_subsec_three}
The numerical results reveal a qualitative transition in the probability of satisfiability in relation to the average edge fidelity across both sub-networks. By distributing the same average fidelity and entanglement generation probability within both networks and varying the average fidelity, we investigate their effects on the overall success rate. For each configuration, the probability of satisfiability is computed by averaging over multiple fidelity distribution scenarios, maintaining fixed average and variance values. Additionally, we analyze the influence of varying the inter-network demand ratio, denoted as \( r \). Higher values of \( r \) indicate a larger proportion of tasks involving inter-network communication, resulting in a more challenging communication environment characterized by increased distances and environmental noise.

We present plots for several configurations: Figures \ref{phase_transition_plot}A and \ref{phase_transition_plot}C depict homogeneous networks, while Figures \ref{phase_transition_plot}B and \ref{phase_transition_plot}D represent networks with a standard deviation of 0.05 in the edge parameters. Figures \ref{phase_transition_plot}A and \ref{phase_transition_plot}B illustrate scenarios in which the probability of satisfiability is related to meeting the fidelity and probability thresholds for quantum key distribution (QKD) tasks, specifically set at 0.9 and 1.6\%, respectively, corresponding to \( R_{\text{eps}} = 10^6 Hz \) and \( R_{\text{sec}} = 10^3 Hz \). Notably, the network sizes considered are identical for both sub-networks. In contrast, Figures \ref{phase_transition_plot}C and \ref{phase_transition_plot}D focus on the probability of satisfiability in relation to achieving the required secure key rate, with variations in average fidelity where we have considered distinct network sizes for both sub-networks.

In homogeneous networks, we observe steep transitions and several jumps in the plot of probability of satisfiability, while the transitions in non-homogeneous networks are comparatively smooth. The various jumps in the homogeneous networks can be explained as follows: the first jump occurs when the intra-network demands begin to satisfy the criteria, while inter-network demands still lag. The second jump arises when the secure key rate in the inter-network demands between the least distant nodes meets the required criteria. This process continues, influenced by the possible graph distances between any two arbitrary nodes for a given network configuration.

All plots are generated for various values of the inter-network demand ratio, \( r \). The average entanglement generation probability is chosen such that it satisfies the required thresholds so that we can see the variation with one parameter. As anticipated, the $P_{SAT}^{(r)}$ decreases as the inter-network demand ratio increases. Additionally, the transition behavior emphasizes the importance of achieving a sufficiently high average fidelity across the network.


\section{Discussion and conclusions}
\label{sec_five}

In this work, we developed an optimization framework to minimize the resource costs involved in constructing connected quantum sub-networks while meeting the threshold required for a given, arbitrary quantum information processing (QIP) task. This optimization approach is illustrated through a simplified entanglement distribution protocol based on entanglement swapping between adjacent edges in the quantum sub-networks. Our analysis builds on a statistical model for quantum networks, as used in Mondal et al. \cite{mondal_ent_topography}, which captures the probabilistic nature of entanglement generation across network edges. This model was based on two important edge parameters—the fidelity of the entangled state over each edge and the probability of successful entanglement generation modeled as random variables with respective probability distribution functions.

We introduced a resource cost function that scales with the number and sizes of the connected sub-networks. This cost was formulated as a non-linear function of the average edge fidelity and the average probability of successful entanglement generation within each sub-network. To solve this constrained non-linear optimization problem, we employed the Lagrangian method in regions where the Karush–Kuhn–Tucker (KKT) conditions are satisfied. This yielded a set of non-linear equations, which can be solved numerically to obtain optimized values for the average edge fidelity and edge generation probability in each sub-network. These optimized parameter values ensured the reliable execution of the specified QIP task across the full quantum network.

Further, we performed numerical simulations to illustrate a transition in the probability of satisfiability of QIP task, where success probability shifts from near-zero to full satisfiability as average edge parameters surpass a certain values. By simulating different levels of inter-network demand, we demonstrated how the ratio of demand in the global network becomes a critical factor in determining the necessary edge parameter values for achieving specified satisfiability probabilities. Additionally, we found that the nature of this transition—whether smooth or abrupt—depends on the network’s configuration, offering insights into how network design impacts resource efficiency and performance. Our approach towards optimising resource requirements for interconnected quantum sub-networks can potentially yield practical targets for average edge parameters for the creation of a modular global quantum network, thereby, advancing the realisation of high-performance, large-scale quantum networks.

\section*{Acknowledgment}
We acknowledge funding from DST, Govt. of India through the SERB grant MTR/2022/000389, IITB TRUST Labs grant DO/2023-SBST002-007, the IITB seed funding and the Chanakya fellowship through the I-HUB Quantum Technology Foundation grant I-HUB/PGF/B-3/2023-24/15.


\end{document}